# Full vibrational characterization of ethylene adsorption on Si(001)-(2x1) by a combined theoretical and experimental approach

Krassimir L. Kostov,[a,b] Rachel Nathaniel,[c] Tzonka Mineva[d] and Wolf Widdra[a]

The vibrational and structural properties of a single-domain Si(001)-(2x1) surface upon ethylene adsorption have been studied by density functional cluster calculations and high-resolution electron energy loss spectroscopy. The detailed analysis of the theoretically and the experimentally determined vibrational frequencies reveals two coexisting adsorbate configurations. The majority ethylene species is di-σ bonded to the two Si atoms of a single Si-Si dimer. The local symmetry of this adsorption complex for ethylene saturation is reduced to $C_2$ as determined by surface selection rules for the vibrational excitation process. The symmetry reduction includes the rotation of the C-C bond around the surface normal and the twist of the methylene groups around the C-C axis. Experimentally 17 ethylene-derived modes are found and assigned for the majority and the minority species based on a comparison with calculated vibrational frequencies. The minority species which can account up to 14 % of the total ethylene coverage is spectroscopically identified for the first time. It is assigned to ethylene molecules di-σ bonded to two adjacent Si-Si dimers (in an end-bridge configuration). One part of these minority species desorbs molecularly at 665 K, about 50 K higher than the majority species, whereas the remaining part dissociates to adsorbed acetylene at temperatures around 630 K. For the latter a di-σ end-bridge like bonding configuration is proposed based on a comparison of the vibrational spectra with data for adsorbed acetylene on Si(100)-(2x1).

## 1. INTRODUCTION

Adsorption of small hydrocarbon molecules at silicon surfaces is an essential aspect of surface chemistry because of its fundamental scientific interest as model system for chemisorption at covalently bound surfaces and the many practical applications based on silicon technologies.[1,2] Despite the impressive number of related studies in the recent literature, there are still controversies even for the seemingly most simple molecules regarding their actual bonding geometry, their vibrational structure, and their reactivity. Here their interaction with the well-characterized and technological important Si(001)-(2x1) surface is of highest relevance. The Si(001)-(2x1) surface offers interesting physical and chemical properties due to the presence of Si-Si dimers within a (2x1) unit cell with two unsaturated (dangling) bonds per dimer.[2] A nominally flat surface consists of two domains with dimer rows oriented along the two equivalent [110] high symmetry directions which are perpendicular to each other. The present study revisits the vibrational structure of chemisorbed ethylene by using a vicinal single-domain Si(001)-(2x1) surface and allows for the first time a full assignment of all symmetry-allowed ethylene-derived vibrations. Additionally a majority and a minority adsorbate species can be identified by comparison with calculated vibrational frequencies based on B3LYP density functional theory (DFT).

Two structural rather different configurations of chemisorbed ethylene on Si(001) have been considered: (i) ethylene adsorption on-top of a single Si-Si dimer by forming two C-Si σ bonds with the Si dimer atoms;[3] (ii) ethylene adsorption between two adjacent silicon dimers by di-σ bonding to Si atoms from neighbouring dimers, named end-bridge configuration.[4] It is commonly accepted now that ethylene chemisorbs in the di-σ on-top configuration.[5-8] In early studies within this bonding model two configurations have been considered with dimer-maintained[5,9-12] and dimer-cleaved structures,[13-16] respectively.

In the first vibrational study using high-resolution electron energy loss spectroscopy (HREELS) Yoshinobu et al.[9] concluded that ethylene is adsorbed nondissociately di-σ bonded to a single dimer on Si(100)-(2x1). The dimer bond retains intact upon adsorption but the symmetry of adsorption complex is reduced, most probably to $C_1$. Experimentally this model has been confirmed by near edge X-ray adsorption fine structure (NEXAFS) and ultraviolet photoemission spectroscopy measurements,[7,17,18] scanning tunnelling microscopy (STM) and photoelectron diffraction study.[19,20] Core level spectroscopy[21-24] showed that the binding energy features, attributed to the buckled atoms of the clean surface Si-Si dimer, almost disappear upon ethylene adsorption. It indicates that at saturation coverage all dimers are involved in the interaction with ethylene molecules which corresponds to one ethylene molecule per dimer (ML).

Besides the di-σ on-top model the existence of the di-σ end-bridge complex was also considered. In combined theoretical and experimental work Marsili et al.[25] found only di-σ on-top adsorbate configuration in the entire coverage region, despite the fact that the di-σ end-bridge configuration was calculated to be energetically more favourable at 1 ML.[26] This discrepancy was explained by kinetic effects which prohibit conversion from the on-top configuration to the end-bridge one at coverages above 0.5 ML. Moreover an energy barrier of 0.12 eV between the precursor state and the end-bridge configuration was calculated whereas for the on-top adsorption this barrier was found almost zero for the incoming molecules at coverages above 0.5 ML.[27] Also using coaxial impact collision ion scattering spectroscopy and computer simulations the adsorption configuration of ethylene on Si(001) surface has been determined to be the di-σ on-top rather than the di-σ end-bridge configuration.[28]

The absolute value of the ethylene saturation coverage on Si(100)-(2x1) had been controversially discussed. Earlier STM studies led to the conclusion that even at higher coverages the ethylene molecules prefer to occupy alternate dimer sites along a dimer row showing nearest neighbour repulsion[8,29-31] and therefore the saturation coverage cannot exceed 0.5 ML. However, now the saturation coverage is accepted to correspond to 1 ML which favours di-σ bonding of a $sp^3$ rehybridized ethylene molecule on each Si-Si dimer.[13,32-34] This discrepancy has been explained by a strong coverage dependence of the adsorption probability which decreases at room temperature significantly above 0.5 ML[21,34] suggesting that the interactions between adsorbates are strong along the dimer rows.[8, 29-30]

The lateral interactions are related to the geometry and symmetry of the adsorbate complex. Experimental evidence for an 1D adsorbate band structure along the dimer rows for the saturated ethylene layer on Si(001)-(2x1) due to direct adsorbate-adsorbate wave function overlap was found based on angle-resolved UV photoemission (ARUPS) and density functional calculations.[6,7] This lateral wave-function overlap is responsible for the observed symmetry reduction of the adsorbate configuration to $C_2$ and a twist of the methylene groups. The distortion was proposed to include an 11° rotation of the molecule C-C axis around the surface normal and a 27° twist of the methylene groups around this axis.[6] Subsequently a polarization-resolved NEXAFS study[35] and resonant inelastic X-ray scattering measurements in combination with density functional calculations[36] confirmed the reduced ($C_2$) symmetry and the distortion of the ethylene adsorbate on the silicon dimer. Recently it was argued using molecular mechanics simulation that steric interactions cause the adsorbed ethylene-dimer rings to be not perpendicular to the surface.[37]

At low temperature (48 K) Nagao et al.[38] observed using HREELS the precursor state for the di-σ bonded ethylene on a Si dimer and identified it as a weakly bonded π-complex adsorbed at the down atom of the buckled Si-Si dimer. Above 70 K the transition to the di-σ chemisorbed ethylene has been observed.

This literature review supports strongly the commonly accepted model of ethylene di-σ bonding on top of a single Si-Si dimer. At saturation coverage the strong lateral interactions between molecules induce a concerted molecular rotation around the surface normal which leads to a $C_{2v}$ to $C_2$ symmetry reduction. This symmetry reduction should lead to the appearance of new vibrational modes which are otherwise symmetry-forbidden. Previous HREELS studies[9,14-16,38,60] which were able to resolve between six and nine adsorbate-induced vibrations did not allow for a conclusive symmetry determination.

This motivated the present study to revisit the full vibrational structure for ethylene chemisorption on single-domain Si(001)-(2x1) with significantly improved energy resolution (2-3 times in comparison to previous works[9,14-16,38,60]) and by direct comparison with the calculated vibrational structure based on density functional theory. With this approach we are able to identify 17 ethylene-derived vibrations. All vibrations are assigned to modes of ethylene majority and ethylene minority species. The former is related to the well-known on-top configuration; however, the vibrational mode assignment has to be corrected with respect to previous studies. For the latter minority species, spectroscopic evidence for end-bridge ethylene adsorption between two Si-Si dimers is presented.

## 2. METHODOLOGICAL ASPECTS

### 2.1 Experimental details

The experiments were carried out in an ultra-high vacuum (UHV) apparatus with a residual gas pressure in the range of $10^{-9}$ Pa. The vicinal Si(001) samples (miscut 5° towards the [110] direction) were cleaned by cycles of $Ar^+$ sputtering followed by heating to 1200 K in UHV and slow cooling with a rate of -2 K/s in order to ensure a low defect concentration.[39] The surfaces exhibited sharp single-domain (2x1) diffraction pattern as monitored by low-energy electron diffraction (LEED). Vibrational spectra were determined by high-resolution electron energy loss spectroscopy (HREELS). The instrumental energy resolution of the spectrometer (Delta 05, VSI) of 8-10 $cm^{-1}$ has been demonstrated on bare and adsorbate covered metal surfaces with corresponding count rate up to $10^6 s^{-1}$.[40-42] However on Si(100), excitations of low-energy plasmons reduce the actual resolution to about 25 $cm^{-1}$ at 100 K as it has been also observed in our earlier study[43] for benzene adsorption on Si(001)-(2x1). But still this resolution is an improvement (by about a factor 2-3) compared to data existing so far.[9,16,38,60] An angle of electron incidence of 60° with respect to the surface normal was used. The azimuth orientation of the sample has been chosen such that the electron scattering plane is spanned by the surface normal and the [110] high symmetry direction. The latter coincides with the Si-Si dimer bond direction of the single-domain Si(001)-(2x1) surface.

Temperature-programmed desorption spectroscopy was performed using a quadrupole mass spectrometer (QMG 112, Balzers) with Feulner cup[44] applying a heating rate of 5 K/s. In more details the experimental setup is described in Ref. 45.

### 2.2 Computational details

The calculations of the vibrational spectra were carried out with Density Functional Theory (DFT) using the hybrid B3LYP exchange correlation functional and two different cluster models. Silicon clusters with two ($Si_{25}H_{38}$) and three ($Si_{21}H_{20}$) Si-Si dimers were cut from the bare silicon surface model which has been optimized with periodic DFT/B3LYP calculations[45].

The three dimer models with ethylene di-σ bonded to a single Si-Si dimer of the Si(001)-(2x1) (on-top configuration) are presented in Fig. 1.

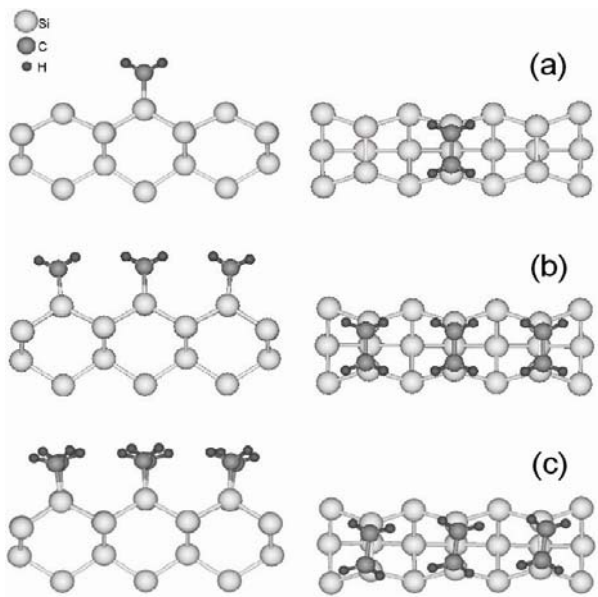

**Fig. 1:** Cluster models for ethylene adsorption on Si(001)-(2x1) that represent different coverages and local symmetries with ethylene di-σ bonding to a single Si-Si dimer: (a) and (b) correspond to $C_{2v}$ symmetry and low and full coverage, respectively. (c) shows the cluster model for $C_2$ symmetry and full coverage. The dangling bonds of the outermost Si atoms in the inner layers are saturated with hydrogen atoms (not shown).

The model in Fig. 1(a) corresponds to a single adsorbed ethylene molecule with $C_{2v}$ local symmetry. The clusters where all dimers are saturated with ethylene in either $C_{2v}$ or $C_2$ symmetry are shown in Figs. 1(b) and (c), respectively. The structural parameters in Fig. 1 were optimized by imposing symmetry constrains as mentioned before. For this purpose the $Si_{21}H_{20}$ clusters were appropriately symmetrised and the Si atoms in the second and third layers were relaxed.

Clusters with two Si-Si dimers were used to model ethylene bridging two adjacent Si-Si dimers by di-σ bonding in end-bridge configuration.

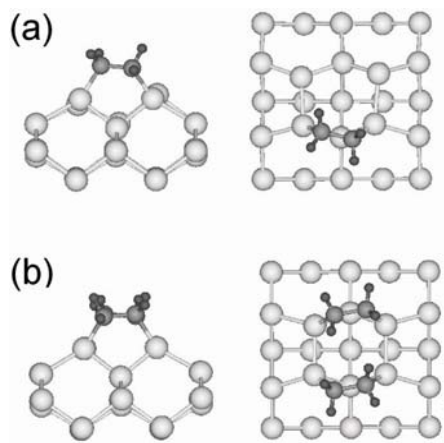

**Fig. 2:** Cluster models for ethylene adsorption on Si(001)-(2x1) with ethylene di-σ bonding between two adjacent Si-Si dimers (end-bridge configuration): (a) and (b) correspond to low and full coverage, respectively. The dangling bonds of the outermost Si atoms in the inner layers are saturated with hydrogen atoms (not shown).

In Fig. 2(a) and (b) the resulting cluster models are displayed for a single ethylene molecule and two ethylene molecules bridging the same dimers, respectively. Geometrical parameters and vibrational frequencies for the two dimer models (Fig. 2) were obtained by relaxing the surface silicon dimers and the $C_2H_4$ molecule(s) while keeping the rest of the Si atoms fixed to their optimized positions from the slab calculations.[45,46]

Geometry optimization and vibrational frequencies were computed with the G03W[47] computer package, using 6-31g[48] basis sets for the dimer silicon atoms, 3-21g basis set[49] for the rest of the cluster Si atoms and 6-31g* bases[48] for the carbon and hydrogen atoms in the $C_2H_4$ molecule(s). The cluster models in Figs. 1 and 2 are validated through comparing their characteristic geometrical parameters with previous theoretical results. Upon adsorption the silicon dimers become symmetric and remain intact as it was also reported from periodic DF studies.[4-6,11,25-27,50] The optimized C-H, C-C and Si-C bond lengths are in the range of 1.10-1.11; 1.52-1.61 and 1.92-1.97 Å, respectively, also agreeing very well with previous periodic DFT results.[5,6,11,17,19,25,27] In the case of the $C_2$ symmetry configuration in Fig. 1(c) the C-C axis is turned to 7.4° around the surface normal resulting in a Si-C-C-Si dihedral angle of 8.9° and in a twist of the methylene groups by 17° with respect to the C-C bonds. These values are in line with previous slab[6] and cluster calculations[36] for $C_2$ symmetry models. It is worth noting that optimizing the paired on-top ethylene adsorbate model on the three dimer clusters without symmetry constrains, led to very similar ethylene rotation around the surface normal as in the model of Fig. 1(c) with $C_2$ local symmetry, e.g. the Si-C-C-Si dihedral angle is 9.4° and the $CH_2$ twist relative to C-C bond is 18.1°. Optimization of singly adsorbed on-top ethylene in $C_2$ symmetry rotated the $C_2H_4$ molecule to the $C_{2v}$ symmetrical configuration.

The expected differences between the computed and experimental frequencies are estimated including also computations for the gas-phase ethylene ($D_{2h}$). The computed frequencies are scaled by 0.963 as suggested for the B3LYP functional and the 6-31g* basis set.[51] In case of gas-phase $C_2H_4$ we obtained asymmetrical ($b_{2u}$ and $b_{1g}$) and symmetrical ($a_g$ and $b_{3u}$) $CH_2$ stretching frequencies (after scaling) higher by 20 – 25 cm$^{-1}$ than the measured ones,[52] while the scissoring, rocking and waging $CH_2$ modes agree with the experimental values within a maximum of 10 cm$^{-1}$. It is found experimentally[9,14-16,38,50] that the asymmetrical $\nu(CH_2)_a$ and symmetrical $\nu(CH_2)_s$ $CH_2$ stretching modes undergo down shifts upon adsorption on Si(001) surfaces in the ranges of 20 - 70 cm$^{-1}$ for $\nu(CH_2)_a$ and of 65 - 80 cm$^{-1}$ for $\nu(CH_2)_s$. We obtained 65-115 cm$^{-1}$ ($\Delta \nu(CH_2)_a$) and 82-100 cm$^{-1}$ ($\Delta \nu(CH_2)_s$). Therefore, an overestimation of the $CH_2$ stretching frequencies up to 40-50 cm$^{-1}$ can be expected.

## 3. RESULTS AND INTERPRETATIONS

### 3.1 Majority ethylene species

For the clean, vicinal single domain Si(001)-(2x1) surface,

phonon features are visible below 550 cm$^{-1}$ as shown in the HREEL spectrum in Fig. 3 (a). Ethylene adsorption at 110 K on this surface leads to the appearance of number of new energy loss peaks in the region between 100 and 1600 cm$^{-1}$ as shown in Figs. 3(b) and (c) as well as a broad loss at about 2930 cm$^{-1}$ in the high-frequency C-H stretch region (Fig. 4). In the Figs. 3 and 4 on- and off-specular data are shown where the scattering plane includes the Si step edges such that the Si-Si dimer bond lies within the scattering (sagittal) plane.

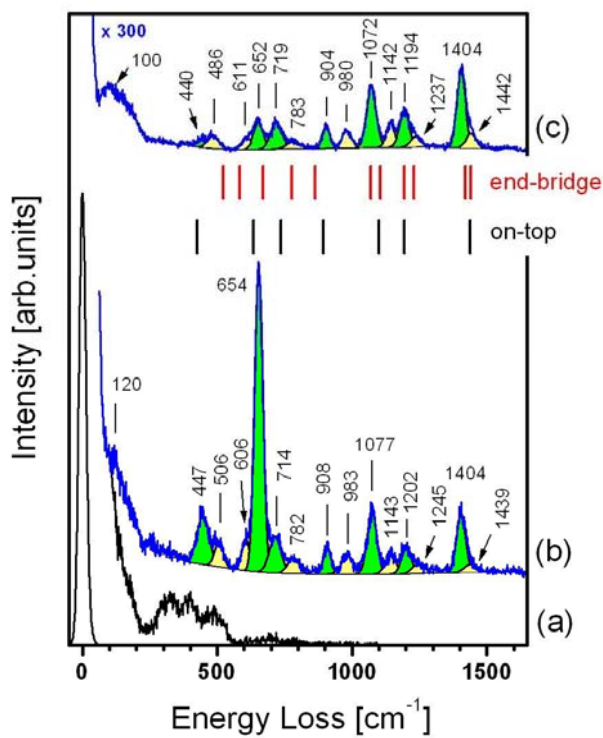

**Fig. 3 (color online):** HREEL spectra of clean Si(001)-(2x1) (a) and upon C$_2$H$_4$ saturation at 110 K (b) in specular scattering geometry. (c) Spectrum as in (b) but in off-specular (ΔΘ=15°) scattering geometry. Calculated vibrational frequencies for ethylene di-σ bonded on-top of a Si-Si dimer and for ethylene di-σ bonded between two adjacent Si-Si dimers (end-bridge configuration) are shown as lower and upper vertical bars, respectively.

In the energy range up to 1600 cm$^{-1}$ thirteen vibrational peaks can be resolved in Fig. 3(b) under specular scattering conditions at ~120, 447, 506, 654, 714, 782, 908, 983, 1077, 1143, 1202, and 1404 cm$^{-1}$ as well as a shoulder at 606 cm$^{-1}$. Besides these losses the deconvolution of the HREEL spectra reveals hints for possible losses at 1245 and 1439 cm$^{-1}$ which explain the asymmetric shape of the energy losses at 1202 and 1404 cm$^{-1}$. As shown in Fig. 3(c) the 1245 and 1439 cm$^{-1}$ loss features can also be identified in off-specular measurements. Most of the energy losses have the same intensities under on- and off-specular scattering conditions. This observation shows the dominant role of the impact scattering mechanism. However the intensities of the 654 and 447 cm$^{-1}$ losses decrease strongly with off-specular angle which indicates dipole scattering for both modes.

In the C-H stretching vibration region which is shown in Fig. 4(a) and (b), a broad and asymmetric peak at 2929 cm$^{-1}$ is observed with a full width at half maximum (FWHM) which is broader than the FWHM of a single well-separated peak (approximately 35 cm$^{-1}$). Under off-specular scattering conditions no significant intensity change is observed in Fig. 4(b). This implies that impact scattering is also the dominant mechanism for excitation of the C-H stretching modes.

The experimentally determined vibrational frequencies are summarized in Tab. 1 and 2 and will be compared with theoretical frequencies based on the B3LYP calculations for different adsorbate structures. We start the comparison with the calculation for ethylene di-σ bonded to the two Si atoms of a single Si-Si dimer. The general accepted model for this adsorption configuration at saturation coverage includes a small C-C axis rotation with respect to the surface normal and a twist of the methylene groups around the C-C bond leading to a reduced C$_2$ symmetry of the adsorbate.[6,7,35,36] Within this symmetry the total-symmetric vibrational eigenmodes (representation A) are dipole- and impact active in specular scattering conditions. In Fig. 3 the calculated frequencies of these modes are indicated as vertical bars for the saturated layer based on the three dimer cluster as depicted in Fig. 1(c). In general a good agreement between the theoretical and experimental frequencies for the most intense losses in Fig. 3 is found. As indicated in Tab. 1 the modes at 1404, 1202, 1077, and 714 cm$^{-1}$ are identified as the dipole-active C-H bending modes, namely the scissoring, rocking, wagging, and twisting motions, respectively. Additionally the vibrations at 908, 654 and 447 cm$^{-1}$ are identified as C-C stretching, symmetric C-Si stretching, and Si-Si dimer stretching modes. The latter is clearly dipole active based on its strongly reduced intensity under off-specular scattering conditions. Based on the good agreement between the experimental and theoretical frequencies with small deviations in the order of 30 cm$^{-1}$ only, the clear assignment of the C-H bending modes has been possible. Note that the resulting assignments of the C-H wagging, the C-H rocking, and the symmetric C-C stretching modes are substantially different from previous work.[9,14-16,38,53] In a recent combined theoretical and experimental work of DiLabio et al.[60] the wagging and C-C stretching frequencies were determined with good agreement with our findings. However substantial differences exist in an assignment of the other modes, for example rocking and twisting vibrations. The comparison between the results of previous experimental studies and the present work is summarized in Table 1. Additionally with the C-H twisting mode assigned, all ethylene-derived modes which are dipole and/or impact active under specular scattering conditions have been determined for the first time here. In the B3LYP calculation it is possible to compare the spectra for C$_{2V}$ and for C$_2$ adsorbate symmetry as well as for ethylene adsorption without neighbouring molecules (noted in Tab. 1 as *single C$_{2v}$*). It is interesting to note that there are only small changes for all three cases within the used cluster model. A clear experimental discrimination between the models as was demonstrated based on angle-resolved UV photoemission[7] is in the present study based only on the values of the vibrational frequencies in Table 1 not possible. The only larger vibrational shift is calculated for the asymmetric C-Si stretch which shifts from 648 to 607 cm$^{-1}$ upon symmetry reduction to C$_2$. But this mode is not symmetry allowed in the specular direction and not assigned here. Despite the small frequency changes in the vibrational spectra computed for the different adsorbate models, these theoretical results become more informative when they are analyzed by accounting also for the surface selection rules. The symmetry reduction from C$_{2v}$ to C$_2$ affects the HREELS surface selection rules: For the chosen scattering plane described in sect. 2.1, a specular analysing direction corresponding to Fig. 3(b) and a C$_{2v}$ symmetric adsorption complex, modes which belong to the A$_2$ and B$_2$ representation are forbidden for dipole as well as impact scattering. On the other hand for a reduced C$_2$ symmetry only the B modes are forbidden. Therefore the visibility of the C-H rocking and the C-H twisting modes at 1202 and 714 cm$^{-1}$ are clear indications of the reduced C$_2$ symmetry. Note that for C$_2$ adsorbate symmetry the B modes are principally allowed in off-specular measurements. However no new peaks are observed in Fig. 3(c) with respect to the specular measurements (Fig. 3b) as they might possess too low intensities as compared with mode being active also in specular direction.

**Tab. 1:** Experimental (Exp.) and theoretical vibrational frequencies in cm$^{-1}$ for di-σ bonded C$_2$H$_4$ on Si(100) in on-top configuration as shown in Fig. 1(a), (b), and (c) for different adsorbate symmetries. The vibrational mode symmetry is indicated. The computed values are scaled by a factor of 0.963 (see text).

| type | Exp. ref. 9 | Exp. ref. 53 | Exp. ref. 16 | Exp. ref. 38 | ref.60 | Exp. this study | single C$_{2v}$ | saturated C$_{2v}$ | saturated C$_2$ | isotope ratio |
|---|---|---|---|---|---|---|---|---|---|---|
| C-H$_2$ stretching, asym. | 3080 | | | 2912 | | | 3010 B2<br>2991 A2 | 3007 B2<br>2988 A2 | 3002 B<br>2984 A | 1.35<br>1.34 |
| C-H$_2$ stretching, sym. | 2955 | 2960 | 2928 | 2905 | | 2928 | 2956 A1<br>2948 B1 | 2961 A1<br>2950 B1 | 2954 A<br>2946 B | 1.37<br>1.37 |
| C-H$_2$ scissoring | 1440 | 1440 | 1416 | 1404 | 1403 | 1404 | 1423 A1<br>1406 B1 | 1439 A1<br>1420 B1 | 1438 A<br>1418 B | 1.33<br>1.36 |
| C-H$_2$ rocking | 900 | | 922 | | 653 | 1202 | 1192 A2<br>998 B2 | 1185 A2<br>985 B2 | 1195 A<br>983 B | 1.33<br>1.39 |
| C-H$_2$ wagging | 1260 | | 1196 | 1182 | 1170<br>1080 | 1077 | 1153 B1<br>1105 A1 | 1156 B1<br>1109 A1 | 1158 B<br>1100 A | 1.22<br>1.46 |
| C-H$_2$ twisting | | | | | 1170<br>1080 | 714 | 734 A2 | 727 A2 | 737 A | 1.20 |
| C-C stretching | 1100 | 1070 | 1076 | 1080 | 887 | 908 | 896 A1 | 893 A1 | 894 A | 0.97 |
| C-Si stretching, asym. (R$_y$) | 800 | | 465 | 452 | | | 651 B1 | 648 B1 | 607 B | 1.02 |
| C-Si stretching, sym. (T$_z$) | 670 | 670 | 655 | 650 | 653 | 654 | 628 A1 | 632 A1 | 634 A | 1.03 |
| R$_x$ | 465 | | | | | | 626 B2 | 627 B2 | 669 B | 1.38 |
| *Si-Si stretching* | | | | | | 447 | 450 A1 | 459 A1 | 425 A | 0.95 |

In addition to the calculation of the vibrations of adsorbed C$_2$H$_4$, computations at the same level of theory were carried out for the deuterated ethylene species as well. The ratio of the scaled B3LYP frequencies of both isotopes were used to determine the isotopic effect for all modes as is summarized in the last columns of Tabs. 1 and 2. This way, calculating an isotope ratio of 0.97, another argument is unequivocally found to attribute the C-C stretching frequency to the observed experimental peak at 908 cm$^{-1}$ (Fig. 1b). Although no HREELS measurements of adsorbed deuterated ethylene were presented here they are not expected to affect the present conclusions. Indeed, considering again the C-C frequency and the calculated isotope ratios, one should expect overlapping of the C-C stretching peak with another intense rocking- and wagging-peaks in the spectral region around 900 cm$^{-1}$. This will prevent the unambiguous determination of C-C stretch peak even with the significantly improved experimental resolution used in our present study. Such overlapping effect might be also responsible for the earlier misassignment of the C-C stretching frequency to above 1000 cm$^{-1}$ based only on the comparison between C2H4- and C2D4- loss spectra.[9,14-16] At that point we can compare the present results with our earlier HREELS/DFT study[45] of adsorbed acetylene on Si(001)-(2x1) for which adsorbate a C=C stretching frequency at 1449 cm-1 was found characteristic for a rehybridization to double carbon-carbon bond. In the present study a further sp$^3$ rehybridization is observed for the adsorbed ethylene in agreement with earlier vibrational studies.[9,16]

In the C-H stretching region a strong energy loss is found at 2928 cm$^{-1}$ which has a weak shoulder at about 2884 cm$^{-1}$ as shown in Fig. 4. Additionally a third even weaker feature is observed at about 2786 cm$^{-1}$. The dominant feature is assigned as the symmetric C-H stretching vibration of the di-σ bonded ethylene on-top of a single dimer. The additional weak features will be assigned to an ethylene minority species.

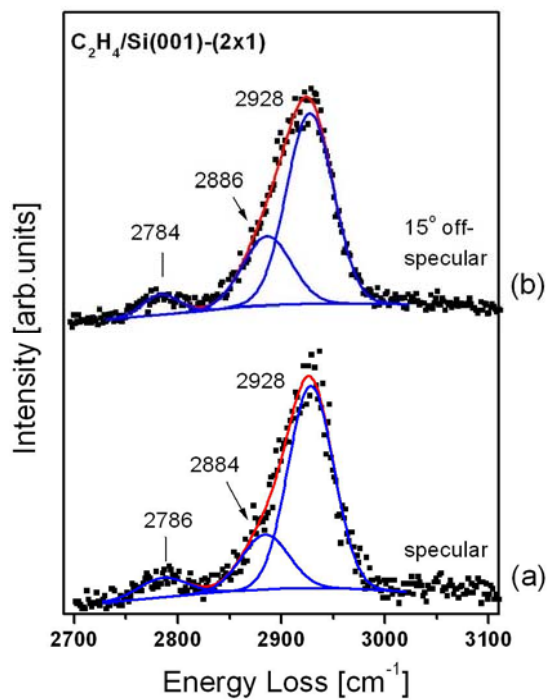

**Fig. 4 (color online):** HREEL spectra in the C-H stretching region for an ethylene saturated Si(001)-(2x1) measured in (a) specular and in (b) off-specular (ΔΘ=15º) directions.

Considering the low-frequency region a loss at ~120 cm$^{-1}$ has been detected in specular scattering geometry in Fig. 3(b) which is clearly visible as a broad structure at about 100 cm$^{-1}$ under off-specular scattering conditions, Fig. 3(c). In this frequency range the frustrated translation and rotation modes of adsorbed ethylene are predicted by theory, for example the dipole and impact active hindered rotation around the surface normal. However we have to note that these low-frequency modes are strongly coupled with clusters vibrations and therefore their unequivocal assignment is not possible.

### 3.2 Ethylene minority species

From the 17 modes which are observed in total above 400 cm$^{-1}$ for the saturated ethylene layer at 110 K (Fig. 3), only the eight dominant vibrations can be assigned to ethylene di-σ bonded on-top of a Si-Si dimer. However, there are weak but well separated additional energy losses at approximately 506, 606, 782, 983, 1143, 1245, and 1439 cm$^{-1}$ visible in Fig. 3(b) and (c). These losses are summarized in Tab. 2 and are assigned to a second minority ethylene species which coexists with the dominant one. Note that simply the number of ethylene-derived peaks already proves that more than one species is present at saturation. In the following we will show that these vibrational modes can be assigned to an ethylene molecule which is di-σ bonded *between* two adjacent Si-Si dimers in an end-bridge configuration as shown in Fig. 2. Fig. 2(a) shows the low-coverage case on a two-dimer cluster model whereas ethylene saturation is illustrated in Fig. 2(b). The calculated vibrational frequencies for these adsorbate configurations are shown in Tab. 2 together with the experimental data and the calculated isotope effect.

In contrast to the on-top species, the symmetry of the adsorbed end-bridge species has been calculated to be reduced to C$_1$. Consequently all modes are allowed by selection rules for both specular and off-specular geometry. According to the calculations (see Tab. 2) the experimentally observed vibrations at 782, 1143, and 1439 cm$^{-1}$ are assigned to the C-H twisting, wagging, and scissoring modes. The losses at 983 and 1245 cm$^{-1}$ are due to the two C-H rocking modes whereas the features at 506 and 606 cm$^{-1}$ are identified as symmetric and asymmetric C-Si stretching vibrations. Note that some losses are seen best in the off-specular data of Fig. 3(c). The comparison between the theoretical and the experimental frequencies reveals again a good agreement. Only for the C-H rocking mode (at 983 cm$^{-1}$) a slightly larger deviation of 86 cm$^{-1}$ is found.

The calculated C-C stretching frequency for end-bridge bonded ethylene is lower compared to that one of on-top ethylene. Whereas for the on-top configuration this frequency has been clearly observed at 908 cm$^{-1}$, the corresponding frequency of the minority end-bridge species which is expected at 864 cm$^{-1}$ is not resolved.

The asymmetrical side of the high-frequency C-H stretching loss at 2929 cm$^{-1}$ in Fig. 4(a) and (b) can be well fitted with a shoulder at 2884 cm$^{-1}$. The latter is interpreted as the symmetric C-H stretch of ethylene in the minority end-bridge configuration. The theoretically calculated symmetrical C-H stretching vibration at 2962 cm$^{-1}$ possesses significantly higher dipole moment and respectively higher intensity than the other symmetrical C-H stretching mode at 2946 cm$^{-1}$ whereas both calculated asymmetrical C-H stretching modes at 3009 and 2987 cm$^{-1}$ consist almost equal dipole moment and intensity, respectively. Therefore the splitting between the frequency of the most intense symmetrical C-H stretch peak and the average asymmetrical C-H frequencies is 36 cm$^{-1}$. The absolute value of these high-lying C-H modes is known to deviate from the experimental ones but the calculated splitting shows a good agreement with the measured value.[45] Therefore we expect the asymmetrical C-H stretching peak to be located at about 2920 cm$^{-1}$ in the experimental spectrum (Fig. 4). Within the C$_1$ symmetry of the end-bridge species this asymmetrical C-H stretching mode is dipole and impact active and should be detectable. However the intense peak at 2928 cm$^{-1}$ of the majority on-top species makes this not visible. Nevertheless we detect a down shift from 2929 cm$^{-1}$ to 2915 cm$^{-1}$ in the annealing experiments to 630 K when the minority of the di-σ end-bridge adsorbed ethylene dominates over the on-top bonded ethylene on the surface due to the intense desorption of the later species at 618 K (see next section).

**Tab. 2:** Experimental (Exp.) and theoretical vibrational frequencies in cm$^{-1}$ for C$_2$H$_4$ on Si(100) in end-bridge configuration as shown in Fig. 2(a) and (b) for single and paired configurations in C$_1$ symmetry. The computed values are scaled by a factor of 0.963 (see text).

| type | Exp. | single C$_1$ | paired C$_1$ | isotope ratio |
|---|---|---|---|---|
| C-H$_2$ stretching, asym |  | 3012<br>2991 | 3009<br>2987 | 1.35<br>1.34 |
| C-H$_2$ stretching, sym | 2884 | 2960<br>2944 | 2962<br>2946 | 1.37<br>1.37 |
| C-H$_2$ scissoring | 1439 | 1429<br>1420 | 1441<br>1421 | 1.35<br>1.36 |
| C-H$_2$ rocking | 1245<br>983 | 1238<br>1062 | 1230<br>1069 | 1.32<br>1.39 |
| C-H$_2$ wagging | 1143 | 1193<br>1090 | 1194<br>1105 | 1.25<br>1.45 |
| C-H$_2$ twisting | 782 | 791 | 778 | 1.23 |
| C-C stretching |  | 872 | 864 | 1.00 |
| C-Si stretching, asym. (R$_x$) | 606 | 564 | 583 | 0.99 |
| C-Si stretching, sym. (T$_z$) | 506 | 523 | 523 | 1.01 |
| R$_y$ |  | 691 | 670 | 1.40 |
| *Si-Si stretching* |  | *383*<br>*367* | *356*<br>*348* | *1.05*<br>*1.06* |

The origin of the weak feature at 2786 cm$^{-1}$ is still not fully clear. Besides been an additional C-H stretch vibration it could also be an anharmonic overtone of the C-H scissoring vibration at 1404 cm$^{-1}$.

### 3.3 Comparison with previous studies

As discussed in sections 3.1 and 3.2, the high resolution spectra allowed identifying two coexisting ethylene adsorption geometries on Si(100) since the experimental energy resolution had been significantly improved. Nevertheless, the dominant vibrational features can be compared with previous studies[9,14-16,38,53,60] as shown in Tab. 1. Whereas larger deviations are observed with the very early studies[9,53] due to reduced sensitivity and energy resolution at that times, the more recent studies[16,38] agree within about 20 cm$^{-1}$ in the frequencies of the dominant features as can be seen in Tab. 1. However, the assignments of the modes are different in comparison to all previous studies. Most prominently, the frequencies of the C-H rocking, wagging and twisting modes are shifted by about 100-300 cm$^{-1}$ (or even more than 500 cm$^{-1}$ compared to the rocking frequency found in the ref. 60) as consequence of the new assignments for the ethylene majority species. Also the important C-C stretching vibration is now assigned to the energy loss at 908 cm$^{-1}$ rather than to a mode in the 1076 – 1100 cm$^{-1}$ range as in the previous experimental studies[9,16,38,53] but in an agreement with the value of 887 cm$^{-1}$ obtained in ref.60. Additionally the Si-Si dimer stretching mode is determined to 447 cm$^{-1}$ for the first time. All the assignments are based on the close agreement with the calculated cluster vibrations for the different ethylene models.

The calculated vibrational frequencies based on the presented DFT cluster calculation can be compared with previous work: Very recently Zhang et al. reported vibrational results based on DFT-GGA calculations using the VASP package.[54] They reported frequencies of 2956, 1388, 1206, and 1079 cm$^{-1}$ for the CH stretching and bending vibrations of di-σ bonded ethylene on-top of a single dimer. These values agree excellently with our data of 2954, 1438, 1195, and 1100 cm$^{-1}$ using the complementary B3LYP cluster approach. Also their C-C and symmetric C-Si stretching frequencies of 920 and 630 cm$^{-1}$ compare well with our calculated data of 894 and 634 cm$^{-1}$, respectively. For the end-bridge ethylene species a fair agreement between the GGA slab and the B3LYP cluster data exist. However the agreement between the experimental data and the B3LYP data seems better: Especially for the CH scissoring and the two C-Si stretching vibrations frequencies of 1441, 583, and 523 cm$^{-1}$ are calculated within the cluster approach as compared to 1401, 669, and 599 cm$^{-1}$ within GGA. Experimentally, these modes are located at 1439, 606, and 506 cm$^{-1}$.

### 3.4 High-temperature adsorbate species

So far the discussion included the structures and the bonding of the majority (on-top of a single dimer) and the minority (end-bridge) ethylene species upon adsorption at a surface temperature of 110 K. The thermal desorption spectrum for the ethylene saturated Si(001)-(2x1) in Fig. 5 exhibits three features: (i) an intense and narrow peak of molecularly desorbing ethylene at about 618 K, (ii) a weak shoulder at higher temperatures, attributed also to molecular ethylene desorption, and additionally (iii) a weak desorption signal from recombining $H_2$ (amu 2) at about 808 K, respectively.

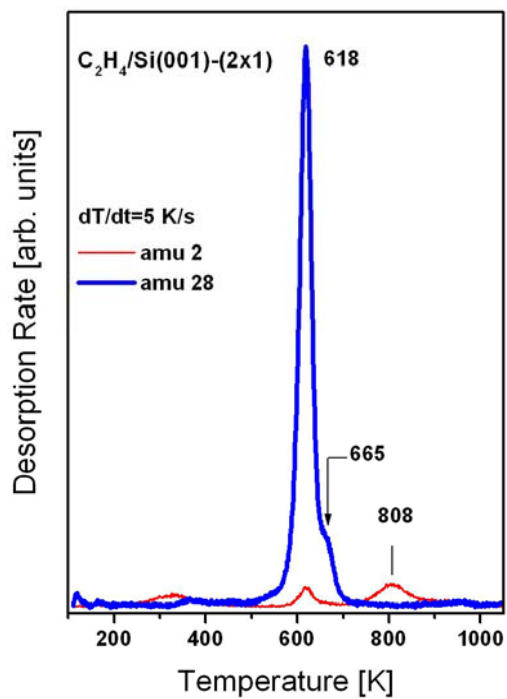

Fig. 5 (color online): Thermal desorption spectra of $C_2H_4$ and $H_2$, respectively, from a saturated ethylene layer on Si(001)-(2x1).

The intense feature at 618 K is characteristic for molecular ethylene desorption from the Si(001) terraces in agreement with previous study[55] whereas the shoulder at 665 K is attributed to desorption dominantly from $C_2H_4$ adsorbed at the step edges of the vicinal Si(001) surface used as well as from ethylene close to defects. Comparing the areas of both thermal desorption peaks at 618 and 665 K we estimate the relative fraction of the second adsorbate species state to approximately 6% with respect to the total amount of desorbing ethylene. An molecular hydrogen desorption peak around 808 K is characteristic for recombinative desorption of hydrogen on Si(001)-(2x1).[56]

The HREEL spectra of the saturated ethylene layer upon subsequent annealing to 320, 500, 630, and 1050 K are displayed in Fig. 6(b)-(f). No changes of the vibrational features are observed upon annealing to 500 K. We conclude that the ethylene layer upon adsorption at 110 K is already in a stable equilibrium structure. No Si-H stretching mode at ~2100 $cm^{-1}$ is visible up to 500 K indicating the absence of ethylene dissociation. However annealing to 630 K causes significant changes of the vibrational spectrum as seen in Fig.

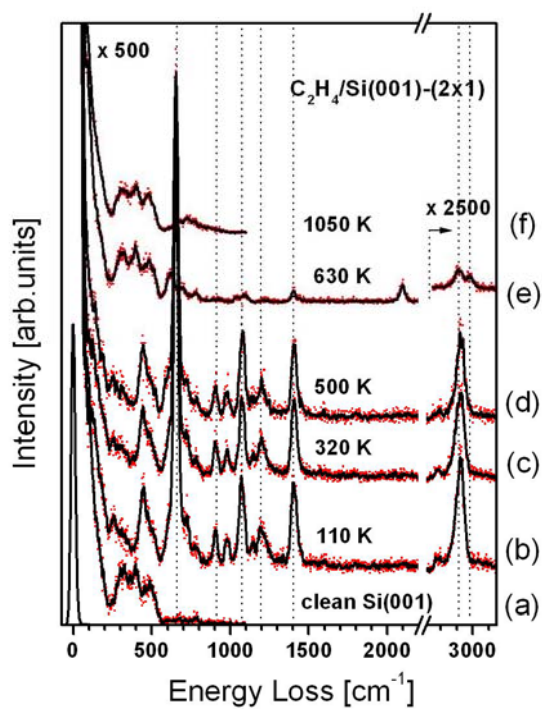

Fig. 6 (color online): HREELS of: (a) the clean Si(001)-(2x1) surface, (b) the ethylene saturated surface upon adsorption at 110 K, (c-f) the same layer as in (b) but annealed to the indicated temperatures.

6(e): A new C-H stretching vibration at 2985 $cm^{-1}$ is observed showing the presence of new adsorbate species. The phonon features below 550 $cm^{-1}$, characteristic for the clean Si(100) surface reappear. Therefore the remaining coverage of this more strongly bonded species is expected to be small. Also a Si-H stretching loss at about 2100 $cm^{-1}$ is observed characteristic for adsorbed hydrogen atoms[57,58] which desorbs molecularly at 808 K. It indicates a small amount of ethylene dissociation besides the well documented molecular desorption (Fig. 5). Upon annealing to 1050 K, only carbon species remain on the surface with a broad loss at about 740 $cm^{-1}$ (Fig. 6(f)).

In order to study the hydrocarbon species responsible for desorption at 665 K, the ethylene saturated surface has been annealed to 630 K. At this temperature the dominant ethylene desorption has occurred and the stronger bonded species still remains on the surface. The thermal desorption spectrum recorded after annealing to 630 K is shown in Fig. 7. Now the high-temperature desorption peak at 675 K dominates which was only visible as shoulder in Fig. 6. Hydrogen desorption is also detected at about 808 K (squares in Fig. 7). This indicates clearly partial ethylene dissociation upon annealing to 630 K. The amount of dissociated ethylene can be determined by comparison of the areas of the hydrogen desorption peak (squares in Fig. 7) relative to the hydrogen desorption signal from a saturated monohydride layer on Si(100)-(2x1) (upper

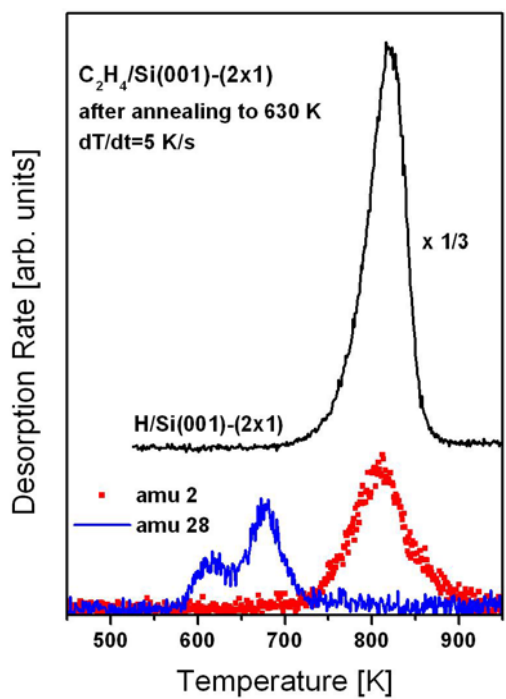
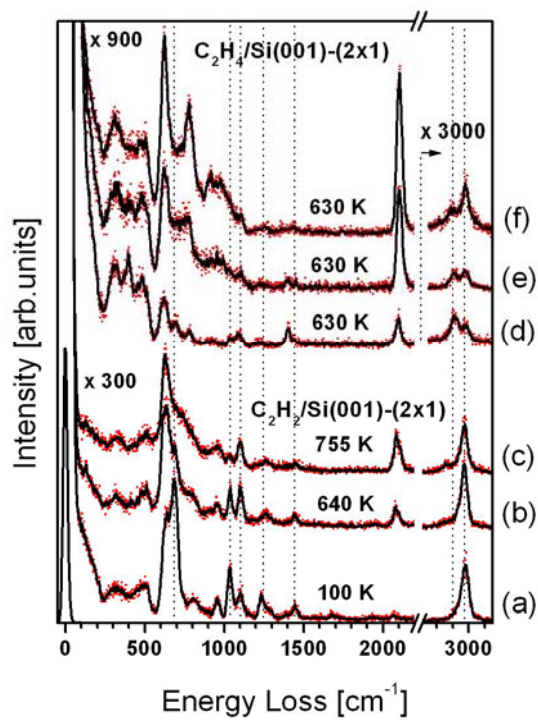

**Fig. 7 (color online):** Thermal desorption spectra of $C_2H_4$ and $H_2$, respectively, from the annealed (at 630 K) saturated ethylene layer. For comparison the $H_2$ desorption spectrum from a saturated monohydride Si(100)-(2x1)-H is shown.

**Fig. 8 (color online):** HREEL spectra for a saturated acetylene layer on Si(001) at 100 K (a) and upon annealing to 640 K (b) and 755 K (c). HREEL spectra for an ethylene saturated layer on Si(001) upon annealing to 630 (d). (e) and (f) show HREEL spectra of the layer (d) but with one (e) and two (f) additional cycles of ethylene saturation at 110 K and annealing to 630 K to enrich the high-temperature species.

solid line in Fig. 7). The latter was prepared using atomic hydrogen exposures on clean Si(100)-(2x1) surface at 650 K.[57] The amount of desorbing $H_2$ molecules from the saturated monohydride layer[59] corresponds to one hydrogen atom per Si surface atom. The comparison of both $H_2$ desorption peaks reveals an ethylene coverage of $0.09 \pm 0.003$ molecules per dimer as the upper limit for the amount of dissociated molecules after annealing to 630 K. To discuss the identity of the hydrocarbon species in Fig. 6 (e) upon annealing to 630 K, HREEL spectra are shown in Fig. 8 (d), (e), and (f) which have been recorded upon three subsequent annealing cycles to 630 K with ethylene saturation at 110 K in between. As reference, the HREEL spectra for a saturated acetylene ($C_2H_2$) layer upon adsorption at 100 K and annealing to 640 and 755 K are shown in Fig. 8 (a), (b), and (c), respectively, as measured on the same vicinal Si(001) sample.

For the ethylene layers which have been annealed to 630 K, Fig. 8 (d)-(f), eleven adsorbate-derived losses at 620, 695, 785, 1040, 1095, ~1225 (broad), 1403, 1440, 2100, 2915 and 2985 $cm^{-1}$ are resolved. The strong losses at 620 and 2100 $cm^{-1}$ can be attributed to adsorbed hydrogen atoms.[57,58] Their presence on the surface is consistent with the hydrogen desorption spectrum in Fig. 7 (square markers) indicating that some dissociation of adsorbed ethylene takes place simultaneously with the intense desorption at 618 K. The measured vibrational frequencies at 785, ~1225, 1440 and 2915 $cm^{-1}$ are in reasonable agreement with the values for the end-bridge bonded ethylene (Tab. 2). Moreover the most intense peak at 654 $cm^{-1}$ characteristic for the on-top adsorbed ethylene (Fig. 3) is absent in Fig. 8 (d)-(f) upon annealing to 630 K. The intense peak at about 620 $cm^{-1}$ is due to contribution of a Si-H bending vibration. Therefore we attribute the thermal desorption peak at 665-675 K to molecular desorption of end-bridge bonded ethylene molecules. However the presence of the Si-H stretch loss at 2100 $cm^{-1}$ indicates that additional to end-bridge adsorbed ethylene there are some hydrocarbon species as a product of ethylene dissociation. Indeed new losses are observed in Fig. 8(d). First of all a new C-H stretching peak which is shifted up in frequency to 2985 $cm^{-1}$ appears. Additionally there are losses at 695 and 1095 $cm^{-1}$ which cannot be attributed to any of the so far considered ethylene configurations (Tabs. 1 and 2).

In the following we will argue that this new species is likely to be adsorbed acetylene which is di-σ bonded between adjacent Si-Si dimers similar to the stronger bound end-bridge ethylene minority species. Reference spectra are presented in Fig. 8 (a)-(c) which show HREELS data upon acetylene adsorption on Si(001)-(2x1) at 100 K and upon annealing to the indicated temperatures. According to our earlier study,[45] acetylene is molecularly adsorbed at 100 K mainly di-σ bonded to a single Si-Si dimer, but also a minority $C_2H_2$ species was found in a di-σ end-bridge like configuration bonded between two adjacent dimers. The corresponding spectrum is shown in Fig. 8(a). Annealing to 640 K and 755 K leads to partial dissociation of adsorbed acetylene, as evidenced by the appearance of Si-H stretch loss at about 2100 cm$^{-1}$, and molecularly adsorbed acetylene (in end-bridge configuration upon annealing to higher temperatures) on the surface.

The losses at 690, 1095 and 2985 cm$^{-1}$ which have been identified as strong acetylene C-H out-of-plane bending, C-H in-plane bending and C-H stretching modes, respectively,[45] in Fig. 8(b) and (c) are also present in the spectra of the annealed ethylene layers in Fig. 8 (d), (e) and (f) and marked by dashed vertical lines. Therefore we conclude that upon annealing a saturated ethylene layer to 630 K, a partial dissociation to adsorbed acetylene and atomic hydrogen occurs. The adsorption configuration of the newly developing acetylene species can be further illuminated based on the frequency of its symmetric C-H in-plane bending vibration. As fingerprint for the acetylene on-top and the acetylene end-bridge configurations, losses at either 1037 or 1104 cm$^{-1}$ have been observed for this mode, respectively.[45] Here similar to the spectrum in Fig. 8(c), the symmetric C-H in-plane bending is observed in Fig. 8(d) and (e) at 1095 cm$^{-1}$ according to end-bridge like acetylene. Therefore we conclude that the adsorbed acetylene produced by the partial dissociation of ethylene at 630 K has a di-σ end-bridge like configuration. It is likely - but without experimental proof so far – that the developing acetylene species resulted from partial dissociation of end-bridge bonded ethylene molecules are adsorbed at step sites of the vicinal Si(001) surface.

Earlier vibrational studies[9,53] suggested that a small amount of adsorbed ethylene dissociate at about 700 K. It was argued that dissociation includes first C-C bond scission and formation of adsorbed $CH_2$ species followed by dehydrogenation at higher temperatures.[9,10] The formation of $CH_2$ species was suggested by the authors of Refs. 9, 10 based on the presence of a low-intensity $CH_2$ scissoring loss at about 1440 cm$^{-1}$. However the C=C stretching vibration for adsorbed acetylene has been identified also in this region, at 1449 cm$^{-1}$, and possesses a low intensity with respect to other acetylene induced peaks.[45] However, here we argue based on the similarities between spectra of the high-temperature species in Fig. 8(d)-(f) with the spectra of adsorbed acetylene in Fig. 8(a)-(c) that the species which is derived from the ethylene dissociation has an adsorbed-acetylene like structure.

Zhang et al.[54] proposed a new reaction channel of $C_2H_2$ with Si(100) surface leading to the formation of "sublayer" adsorbed acetylene bonded to a sublayer and a dimer silicon atom. This intermediate was found to be stable at low (150 K) temperature and to convert to the more stable "normal" interdimer end-bridge acetylene at room temperature. In this theoretical study[54] a large C-H stretch splitting (~100 cm$^{-1}$) was calculated for the sublayer acetylene. Such a splitting is not observed in the present study for the acetylene resulting from the ethylene dissociation. Indeed, by increasing the surface concentration of this dissociation product only one broad C-H stretch peak grows up at 2985 cm$^{-1}$ (Fig. 8(d)-(f)) (the low-intense C-H stretch peak at 2915 cm$^{-1}$ was unequivocally attributed to remaining end-bridge ethylene). As mentioned above this is in agreement with our earlier HREELS/DFT study of acetylene adsorption on Si(001)-(2x1)[45] where such a splitting in the experimental spectrum of adsorbed acetylene was not observed even at 100 K. Therefore the formation of sublayer acetylene as a result of the ethylene decomposition at significantly higher temperature (>600 K) seems to be not likely. In more details this issue will be considered in a forthcoming paper.

Based on the TPD and HREELS results, we estimate roughly the concentrations of the low-temperature di-σ bonded on-top and end-bridge like ethylene species upon adsorption. As discussed before the coverage of the acetylene which is formed at elevated temperature based on ethylene dissociation been estimated to at most 0.09 ML. Associating the higher temperature thermal desorption feature at 665 K with end-bridge like ethylene, the ratio of the areas of both molecular ethylene desorption peaks at 618 and 665 K of 94:6 results in a coverage of 0.05 ML for end-bridge like ethylene. Together with the 0.09 ML acetylene as discussed before, an upper limit of the total end-bridge ethylene coverage can be given by 0.14 ML. Therefore a coverage for on-top ethylene of at least 0.86 ML follows.

## 4. CONCLUSIONS

This study presents a full vibrational characterization for ethylene adsorption on a single-domain Si(001)-(2x1) surface based on B3LYP density functional theory and high-resolution electron energy loss spectroscopy. The detailed analysis of theoretically and experimentally determined frequencies and their comparison reveal two different but coexisting bonding configurations of chemisorbed ethylene upon adsorption at and below room temperature: A majority species (approximately 0.86 molecules per dimer at saturation) which is di-σ bonded on-top of a single Si-Si dimer and a minority (~ 0.14 molecules per dimer) which is di-σ bonded to two adjacent Si-Si dimers in an end-bridge configuration.

The majority ethylene species which adsorbs on defect-free Si(001) terraces desorbs molecularly at 618 K. For a cluster model with a single ethylene molecule di-σ bonded on-top of a Si-Si dimer, DFT calculations favour (representative for low adsorbate coverage below 0.5 ML) a $C_{2v}$ symmetry configuration even if initially $C_2$ symmetry constrains were imposed. For the saturated chemisorbed layer the symmetry is determined experimentally and theoretically to $C_2$ as found previously for slab calculations.[7] The C-C axis is turned by 7.4° around the surface normal and the methylene groups are twisted by 17° with respect to the C-C bond.

For the first time all dipole- and impact-active vibrational modes within the specular scattering geometry were detected and assigned for the majority ethylene species. Especially the C-C stretching and the $CH_2$ twisting frequency are found at 907 and 714 cm$^{-1}$, respectively. Additionally the Si-Si dimer stretching frequency is observed at 447 cm$^{-1}$ in good agreement with theory. A vibrational identification of the minority species as end-bridge like ethylene is presented. Upon annealing about one third of the minority species desorbs molecularly at 618 K, whereas about two thirds dissociate to chemisorbed acetylene. For the latter a di-σ end-bridge like bonding configuration is proposed based on a comparison of the vibrational spectra with data for acetylene on Si(100)-(2x1).

## Acknowledgments

This work was partially funded by a Grand No. X-1316 of the Ministry of Education and Science, Bulgaria and by the Deutsche Forschungsgemeinschaft (DFG).

[a] Institute of Physics, Martin-Luther-University Halle-Wittenberg, 06099 Halle, Germany.
Fax: ++49345527160; Tel: ++49345525560; E-mail: wolf.widdra@physik.uni-halle.de

[b] Institute of General and Inorganic Chemistry, Bulgarian Academy of Sciences, 1113 Sofia, Bulgaria.
Fax: ++35928705024; Tel: ++35929792536; E-mail: klkostov@bas.bg

[c] Institute of Catalysis, Bulgarian Academy of Sciences, 1113 Sofia, Bulgaria.
Fax:++ 3592712967; Tel:++ 35929792550; E-mail: rnath@ic.bas.bg

[d] UMR 5253 CNRS/ENSCM/UM2/UM1, Institut Charles Gerhardt Montpellier, 8 rue de 1' Ecole Normale, 34296 Montpellier Cédex 5, France
Fax:++ 33467163470; Tel:++ 33467163468; E-mail: tzonka.mineva@enscm.fr